# First Terrestrial Soft X-ray Auroral Observation by
# The Chandra X-ray Observatory

Anil Bhardwaj[1,*], G. Randall Gladstone[2], Ronald F. Elsner[3], Nikolai Østgaard[4]
J. Hunter Waite, Jr.[5], Thomas E. Cravens[6], Shen-Wu Chang[7],
Tariq Majeed[5,9], and Albert E. Metzger[8]

[1] *Space Physics Laboratory, Vikram Sarabhai Space Centre, Trivandrum 695022, India*
[2] *Department of Space Science, Southwest Research Institute, San Antonio, TX 78228, USA*
[3] *NASA Marshall Space Flight Center, Space Science Branch, NSSTC/XD12, Huntsville, AL 35805, USA*
[4] *Department of Physics and Technology, University of Bergen, Bergen N-5007, Norway*
[5] *Department of Atmospheric, Oceanic, & Space Sciences, University of Michigan, Ann Arbor, MI 48109, USA*
[6] *Department of Physics & Astronomy, University of Kansas, Lawrence, KS 66045, USA*
[7] *University of Alabama in Huntsville, NSSTC, XD12, Huntsville, AL 35805, USA*
[8] *Jet Propulsion Laboratory, Pasadena, CA 91109, USA*
[9] *Now at Department of Physics, American University, Sharjah, United Arab Emirates*

*Corresponding author: Tel +91-471-2562330; fax +91-471-2706535; email:
Anil_Bhardwaj@vssc.gov.in

<u>See also Press Release by NASA (</u>News release: 05-192) <u> and CXC (</u>RELEASE: 05-10)

<u>on this paper  :</u>

**"Chandra Looks Back At The Earth" - December 28, 2005**

**http://chandra.harvard.edu/press/05_releases/press_122805.html**

**http://chandra.harvard.edu/photo/2005/earth/**

**Abstract**

Northern auroral regions of Earth were imaged with energetic photons in the 0.1-10 keV range using the High-Resolution Camera (HRC-I) aboard the Chandra X-ray Observatory at 10 epochs (each ~20 min duration) between mid-December 2003 and mid-April 2004. These observations aimed at searching for Earth's soft (<2 keV) X-ray aurora in a comparative study with Jupiter's X-ray aurora, where a pulsating X-ray "hot-spot" has been previously observed by Chandra. The first Chandra soft X-ray observations of Earth's aurora show that it is highly variable (intense arcs, multiple arcs, diffuse patches, at times absent). In at least one of the observations an isolated blob of emission is observed near the expected cusp location. A fortuitous overflight of DMSP satellite F13 provided SSJ/4 energetic particle measurements above a bright arc seen by Chandra on 24 January 2004, 20:01–20:22 UT. A model of the emissions expected strongly suggests that the observed soft X-ray signal is bremsstrahlung and characteristic K-shell line emissions of nitrogen and oxygen in the atmosphere produced by electrons.

## 1. Introduction

Both Earth and Jupiter are magnetic planets having dense atmospheres and well-developed ionospheres and magnetospheres (e.g., Schunk and Nagy, 2004; Bagenal et al., 2004). Thus, several atmospheric, ionospheric and magnetospheric phenomena are found to exist on these planets that are similar in nature, including auroral processes and emissions (e.g., see reviews by Bhardwaj and Gladstone, 2000; Waite and Lummerzheim, 2002; Galand and Chakrabarti, 2002; Clarke et al., 2004). Both planets also exhibit X-ray emission associated with their auroras and their non-auroral disks (Stadsnes et al., 1997; Petrinec et al., 2000; Bhardwaj et al., 2002, 2006b; Waite and Lummerzheim, 2002; Bhardwaj, 2006). It is well known that the X-ray aurora on Earth is generated by energetic electron bremsstrahlung (e.g., Berger and Seltzer, 1972; Stadnes et al., 1997; Petrinec et al., 2000; Bhardwaj et al., 2006b), and the X-ray spectrum of the aurora has been very useful in studying the characteristics of energetic electron precipitation (Stadnes et al., 1997; Østgaard et al., 1999, 2001). On the other hand, auroral X-rays from Jupiter are mainly produced by charge-exchange of highly-ionized energetic heavy ions precipitating from the outer magnetosphere and/or solar wind (Gladstone et al., 2002; Cravens et al., 1995, 2003; Branduardi-Raymont et al., 2004, 2005, 2006b; Elsner et al., 2005; Bhardwaj et al., 2006b; see review by Bhardwaj and Gladstone, 2000 for earlier studies). Disk X-ray emission from both planets is largely due to scattering and fluorescence of solar X-rays (McKenzie et al., 1982; Petrinec et al., 2000; Maurellis et al., 2000; Bhardwaj et al., 2005, 2006a, 2006b).

Recent X-ray observations of Jupiter by Chandra X-ray Observatory (CXO) have demonstrated that most of Jupiter's northern auroral X-rays come from a 'hot spot' located poleward of the main ultraviolet auroral oval, thus pointing to a particle source population in the outer magnetosphere (Gladstone et al., 2002; Elsner et al., 2005). Interestingly, the hot spot X-rays pulsate with an approximately 45 (±20) minute period, a period similar to that reported for high-latitude radio and energetic electron bursts observed by near-Jupiter spacecraft (cf. MacDowall et al, 1993; McKibben et al., 1993; Elsner et al., 2005). One possible explanation for the X-ray hot spot is high-latitude reconnection of interplanetary magnetic field (IMF) lines, allowing the precipitation of heavy solar wind ions into Jupiter's cusp region. Since the solar wind heavy ions are highly ionized they can produce soft X-rays by charge exchange and excitation (i.e., the same mechanism responsible for cometary X-rays, cf., Cravens, 2002) as they encounter Jupiter's upper atmosphere.

The identical process should operate at Earth as well. However, while hard X-ray emissions from electron bremsstrahlung are well known in the terrestrial aurora (i.e., Stadsnes et al., 1997; Østgaard et al., 2001), surprisingly, there have been no dedicated searches for auroral X-ray emissions at energies <2 keV. A few limb scans of the nighttime Earth at low latitude by the X-ray astronomy satellite, HEAO-1, in the energy range 0.15 keV to 3 keV, showed clear evidence of the K-α lines of Nitrogen and Oxygen sitting on top of the bremsstrahlung spectrum (Luhmann et al., 1979). Dedicated auroral X-ray experiments have not measured X-rays below 2-3 keV, e.g., the PIXIE X-ray imager on the Polar spacecraft measured X-rays in the range 3-60 keV (Imhof et al., 1995). The high apogee of the Polar satellite (~9 $R_E$) enabled PIXIE to image the entire auroral oval with a spatial resolution of ~700 km. PIXIE data showed that the substorm X-rays brighten up in the midnight sector and have a prolonged and delayed maximum in the morning sector due to the scattering of eastward-drifting electrons (Østgaard et al., 1999). Statistically, the X-ray bremsstrahlung intensity is largest in the midnight substorm onset, is significant in the morning sector, and has a minimum in the early dusk sector (Petrinec et al., 2000). During the onset/expansion phase of a typical substorm the electron energy deposition power is about 60-90 GW, which produces around 10-30 MW of bremsstrahlung X-rays (Østgaard et al., 2002).

We have conducted a series of short duration CXO observation of Earth aimed at searching for soft (<2.0 keV) X-ray auroral emissions for a comparative study with Jupiter's X-ray aurora. In this paper we report the first observation of soft (0.1-10 keV) X-ray emission from terrestrial aurora and present some of the preliminary results.

## 2. Chandra X-ray Observation of Aurora

Ten HRC-I (High Resolution Camera in imaging mode) observations were performed when CXO was near apogee (~20 $R_E$; Chandra moves in a highly elliptical 63.5 hr orbit) and timed during northern winter of 2003-2004, so that the northern polar region was mostly dark and solar fluoresced X-ray contamination could be avoided (cf. Table 1). HRC-I is sensitive to X-rays in the 0.1-10 keV band (with a peak efficiency near ~0.6-2.0 keV) and has spatial resolution of 0.5″ (~0.3 km at Earth from Chandra apogee). Each observation was ~20 min in duration (cf. Table 1), with fixed pointing in right ascension and declination, arranged so that the parallax motion of the Earth allowed the north polar cusp to drift through the HRC-I field-of-view (30′ × 30′, or about 1150 km × 1150 km at the Earth as seen from Chandra apogee) at a variety of local times.

The time-tagged photon list data were analyzed with the CIAO software (v3.1) available from CXO (http://cxc.harvard.edu/ciao/). Each observation was corrected for the Earth's motion and converted into brightnesses images using appropriate exposure maps. As the mean photon energy is not known, a typical HRC-I effective area of 40 cm$^2$ was assumed in converting counts to Rayleighs.

These CXO-HRC-I observations (combined with a test observation taken on February 7, 2003, cf., Table 1) reveal a range of morphologies of X-ray auroras in the northern polar region of the Earth. These first Chandra soft (0.1-2 keV) X-ray observations of Earth's aurora show that it is highly variable; they appear sometime as intense single arcs, other times as multiple arcs, or as diffuse patches, and at times almost absent. Figure 1 shows six examples of these different manifestations of the soft X-ray aurora. Generally, auroral X-rays are brighter when Bz is negative for few hours prior to the Chandra observation. In one of the observations an isolated blob of emission is observed (cf. Figure 1e) near the location where we expect the cusp to be, perhaps giving an indication of solar wind charge-exchange signature in X-rays. However, this identification is still tentative.

Figure 2 shows the CXO observation on 24 January 2004, 20:01–20:22 UT, when a bright arc is seen. Unfortunately no good view of the aurora from IMAGE/FUV is available at this time. TIMED/GUVI provided only a marginal view of this area about 20 min prior to the Chandra observation. However, the DMSP F13 flew over this region at the same time that the HRC-I observed the bright arc. Taking advantage of this situation, in the following, we calculate the X-ray emissions expected from bremsstrahlung and characteristic K-shell line emissions of nitrogen and oxygen in the atmosphere produced by the electrons measured by the DMSP F13 SSJ/4 electrostatic analyzers and compare it with our HRC-I observation.

## 3. DMSP Observation and Electron Bremsstrahlung Model

The DMSP satellites are sun-synchronous polar orbiting satellites with orbital period of 101 min and a nominal altitude of 830 km. The satellites are three-axis stabilized, and the detector always points toward local zenith. The SSJ/4 electrostatic analyzers on board the DMSP satellites measure electrons and ions from 32 eV to 30 keV in 19 logarithmically spaced steps (Hardy et al., 1984). One complete electron and ion spectrum is obtained every second. At the latitudes of interest in this paper, this means that only particles at pitch angles <15°, well within the atmospheric loss cone, are observed. To estimate the X-ray production we have used 10 s averaged spectra from the SSJ/4.

In the energy range 0.1–2.0 keV there are two components to the X-ray spectrum produced by electrons; the continuous bremsstrahlung spectrum and the strong K-α emission lines from nitrogen (at 0.393 keV) and oxygen (at 0.524 keV).

To compute the X-ray bremsstrahlung production from electron precipitation, we use a look-up table of the angular dependent X-ray spectra produced by single exponential electron spectra. The look-up table is generated on the basis of the 'general electron-photon transport code' of Lorence (1992). This code takes into account the scattering of electrons, production of secondary electrons, angular dependent X-ray production, photoelectric absorption of X-rays and Compton scattering of X-rays.

From previous studies of X-ray measurements it has been found that a sum of two exponentials can often be used to represent the X-ray spectra very well [e.g., Goldberg et al., 1982]. From a rocket experiment in the post midnight sector during the recovery phase of a substorm (Østgaard et al., 1998), when both X-ray measurements and electron measurements were available, it was found that both the electron spectrum and the X-ray spectrum could be represented by a sum of two exponentials, giving very good correlation when comparing measured and calculated electron spectra and measured and calculated X-ray spectra. Figure 3 shows two-exponential fits to eight 10-s averaged spectra measured by DMSP F13 (1959:55–2001:04 UT) as the satellite passed through the arc observed by the Chandra. Using 15 of the 19 channels of the SSJ/4 detector we obtain electron spectra from 0.1 to 30 keV. The double exponentials represent the measurements fairly well, but the fit may be worse for energies above 30 keV. The existence of a hard tail in the electron spectrum above 30 keV would lead to an underestimate of the X-ray production. In some of the spectra such a hard tail can be seen, but as the flux level falls off rapidly at the high energies, we do not expect the contribution to the X-ray production to be very large. Keeping in mind that our approach may lead to a slight underestimate of the predicted X-ray bremsstrahlung, we conclude that the double exponential fit can be used to represent the electron measurements from the DMSP satellite in the present study. A similar approach and conclusion was obtained by Østgaard et al. (2000) when comparing estimated and observed X-rays from DMSP and PIXIE, respectively, and we refer to that paper for further documentation on the methodology used in the present study.

To estimate the production of the K-α emission lines we have used the results from Luhmann and Blake (1977), where they have given the relative production of line emissions and bremsstrahlung in a 100-eV band around these lines (cf. Figure 4 of Luhmann and Blake, 1977). Folding these results with the DMSP spectra (Fig. 3) measured over the X-ray arc, we find that the bremsstrahlung X-rays must be multiplied by a factor of 4 to 6, depending on the shape of the electron spectrum.

## 4. Discussion and Summary

Figure 4 shows the time series of the calculated X-ray fluxes in the two energy ranges of 0.1–2.0 keV (middle panel) and 2.0–10 keV (bottom panel) calculated using the DMSP F13 measured electron fluxes (shown in the top panel) on January 24, 2004. Most of the X-rays (>~70%) are produced in the softer 0.1–2.0 keV X-ray band. The X-rays in the 0.1-2.0 keV band include both bremsstrahlung and characteristic line emissions of the atmosphere. In a narrow energy band of ~100-eV around 0.4-0.5 keV, the K-α oxygen and nitrogen line emissions are 30-50 times stronger than the bremsstrahlung. The X-ray brightness from the brighter part of the arc seen in the HRC-I image (cf. Figure 2) is estimated to be ~1 × 10$^5$ photons cm$^{-2}$ sr$^{-1}$ s$^{-1}$. From Figure 4 we find that the energy-integrated bremsstrahlung X-ray flux over the 0.1–10 keV range is likewise about 1 × 10$^5$ photons cm$^{-2}$ sr$^{-1}$ s$^{-1}$ at the peak. This strongly suggests that the Chandra HRC-I observed X-ray emissions on January 24, 2004 are X-rays (bremsstrahlung and line emissions) from electron precipitation. Since the DMSP F13 satellite is not going through the cusp region and the X-ray arc is basically in the post-noon/dusk sector (and not in the cusp) – this is a very reasonable result.

It is interesting to note that most of the auroral X-rays at Jupiter are produced due to charge exchange collision between precipitating highly-ionized heavy (O, S and/or C) energetic (>1 MeV) ions and ambient neutrals in the polar upper atmosphere. These X-rays are line emissions and arise as the heavy ions are nearly stripped of electrons while precipitating, and then (through further collisions) are either directly excited or charge exchanged into an excited state, which emits an X-ray photon upon decay back to the ground state (cf. Cravens et al., 2003; Elsner et al., 2005; Bhardwaj et al., 2006b). Earth auroral X-rays are known to be produced due to bremsstrahlung (and line emissions (around 0.4-0.5 keV) at lower energies) by precipitating electrons, and the present study suggests that even at softer (<2.0 keV) X-ray energies the source appears to be the same. The spectrum of X-rays at these soft energies would help to further strengthen this statement. The other possible source of Earth auroral X-rays could be the precipitation of solar wind ions in the cusp region and production of X-rays in charge exchange collision of highly-ionized heavy (C, N, O) solar wind ions with ambient atmospheric species. The polar cusps are extremely dynamic regions that play a key role in the transfer of mass, momentum, and energy from the solar wind into the magnetosphere and upper atmosphere systems. It has been argued that cusp is an important source region of energetic particles (Fritz, 2001; Chen and Fritz, 2005). Recently, it has also been suggested that the solar wind ions can be energized at the Earth's bow shock before entering the cusp (Chang et al., 1998, 2000). For high charge state ions ($O^{+6}$), the energy can be as high as 1 MeV (Chang et al., 1998, 2000). This is important, as observations at Jupiter suggest that even solar wind ions need to be accelerated if they are to provide the observed power of the auroral X-rays (Cravens et al., 2003; Elsner et al., 2005). In 10 April 2004 Chandra Earth observation we do see a spot of X-ray emission near (though not very close) the calculated location of cusp (cf. Figure 1e). However, in the absence of an X-ray spectrum of this emission-blob, and the lack of proper ion flux measurements, it is difficult to decipher the possible source mechanism. We will propose to conduct observations in the future using Advanced CCD Imaging Spectrometer instrument on Chandra or with XMM-Newton that both have spectral resolution to distinguish between line emissions from highly-charged ions and continuum (cf. Elsner et al., 2005; Branduardi-Raymont et al., 2005, 2006).

**Acknowledgements**

This work is based on observations obtained with Chandra X-ray Observatory and was supported by grant from the Chandra X-ray Center. A part of this research was performed while A. Bhardwaj held a National Research Council Senior Resident Research Associateship at NASA Marshall Space Flight Center. We acknowledge the excellent help we received from Michael Juda of CfA in performing these Earth X-ray observations with Chandra. We also acknowledge help from Johan Stadsnes in locating references related to early studies on Earth X-rays.

**Table 1.** Log of Observation by the Chandra X-ray Observatory

| Date | Chandra OBSID | Start time value[1] | Stop time value[1] | IMF By, Bz[2] (nT) | Dipole Tilt[3] (°) |
|---|---|---|---|---|---|
| 07 February 2003 | 4410 | 01:24 | 01:42 | -3.4, -2.5 | -21.7 |
| 16 December 2003 | 4456 | 04:14 | 04:35 | 1.3, 1.6 | -33.7 |
| 06 January 2004 | 4457 | 07:57 | 08:18 | 0.6, -1.8 | -29.4 |
| 24 January 2004 | 4459 | 20:01 | 20:22 | -4.9, -3.8 | -12.2 |
| 30 January 2004 | 4460 | 03:06 | 03:26 | 7.2, 2.4 | -27.1 |
| 14-15 February 2004 | 4461 | 23:58 | 00:18 | -2.5, -1.4 | -15.8 |
| 28 February 2004 | 4458 | 05:01 | 05:22 | 8.6, -2.4 | -18.6 |
| 04 March 2004 | 4462 | 12:04 | 12:32 | 2.3, 0.4 | -2.7 |
| 07 March 2004 | 4463 | 03:32 | 03:53 | 2.7, 0.7 | -14.9 |
| 10 April 2004 | 4464 | 12:28 | 12:48 | 1.3, 2.0 | +12.9 |
| 13 April 2004 | 4465 | 03:46 | 04:07 | -2.7, -1.2 | -0.9 |

[1] hh:mm in UT.

[2] The IMF (interplanetary magnetic field) By and Bz are in GSM (Geocentric Solar Magnetospheric System) coordinates taken from the OMNIWeb http://nssdc.gsfc.nasa.gov/omniweb/form/dx1.html, and interpolated to Chandra observation times from the hourly averaged-values.

[3] Dipole tilt is defined as the angle between the Earth's north dipole axis and the GSM z-axis (cf. Zhou et al., 1999). This angle is positive when the dipole tilts towards the Sun, and negative when the dipole tilts away from the Sun.

**(a)**

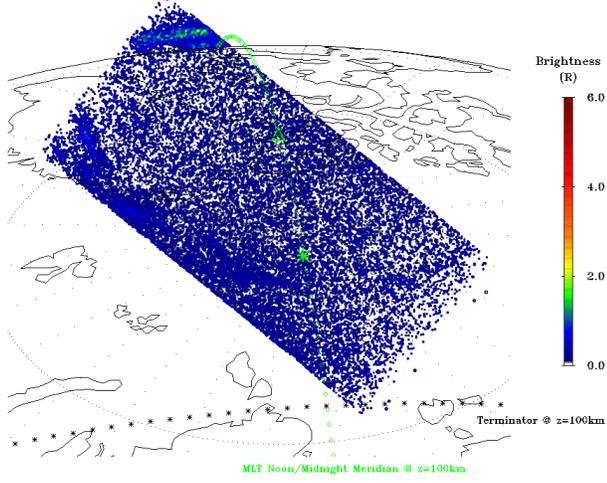

**(b)**

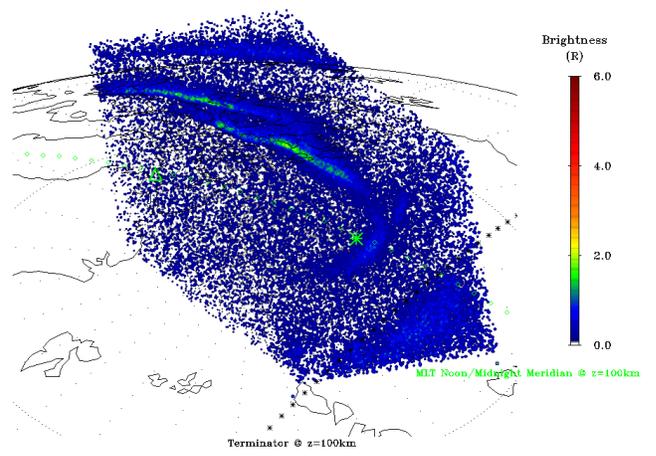

**(c)**

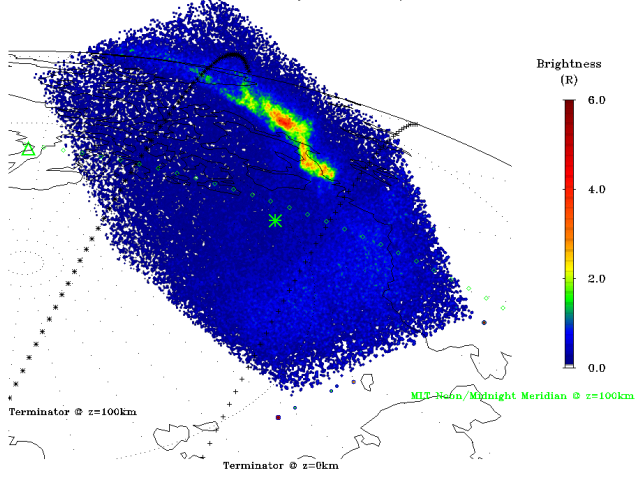

**(d)**

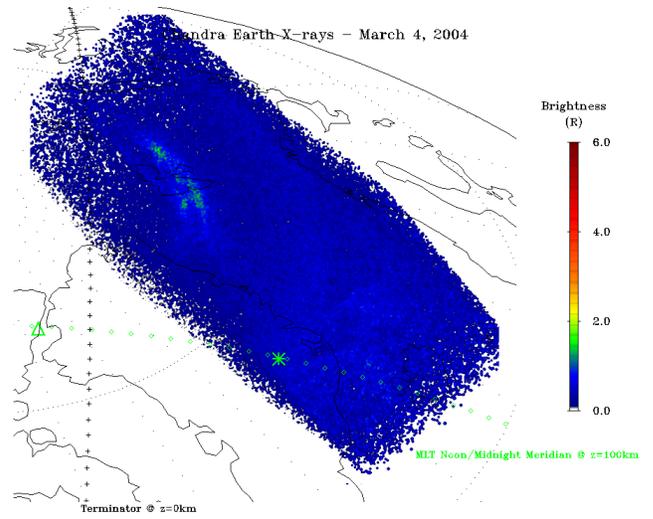

**(e)**

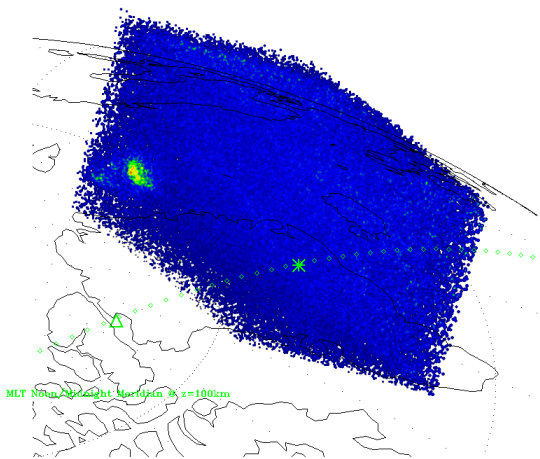

**(f)**

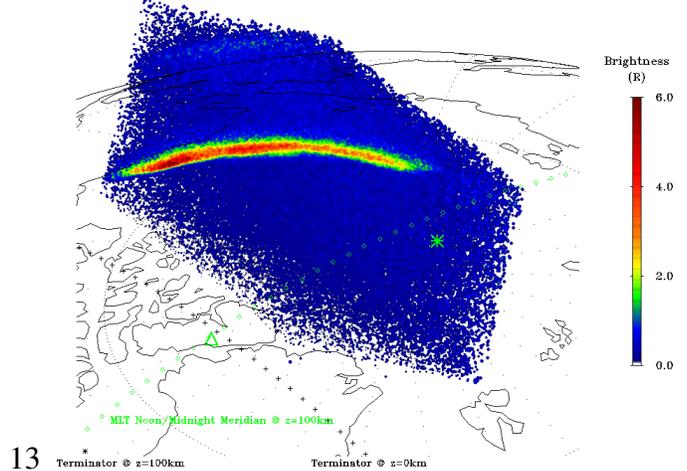

**Figure 1**.  Six example X-rays images (shown on the same brightness scale) of the north polar region obtained by Chandra HRC-I on different days (marked on top of mages), showing large variability in soft (0.1-10.0 keV) X-ray emissions from Earth's aurora. Note that the images are not snap shots, but are ~20-min scans of the northern auroral region in the HRC-I field-of-view. The brightness scale in Rayleighs (R) assumes an average effective area of 40 cm$^2$. 1 R = 10$^6$ photons cm$^{-2}$ s$^{-1}$ (4$\pi$sr)$^{-1}$. Black crosses and asterisks mark the day-night terminator (shadow boundary) at an altitude of 0 and 100 km, respectively. Green diamonds indicate the noon/midnight MLT meridian at an altitude of 100 km. A green triangle marks the magnetic pole and a green asterisk marks the expected cusp location. The location of cusp is calculated using relations given in Newell et al. (1989).

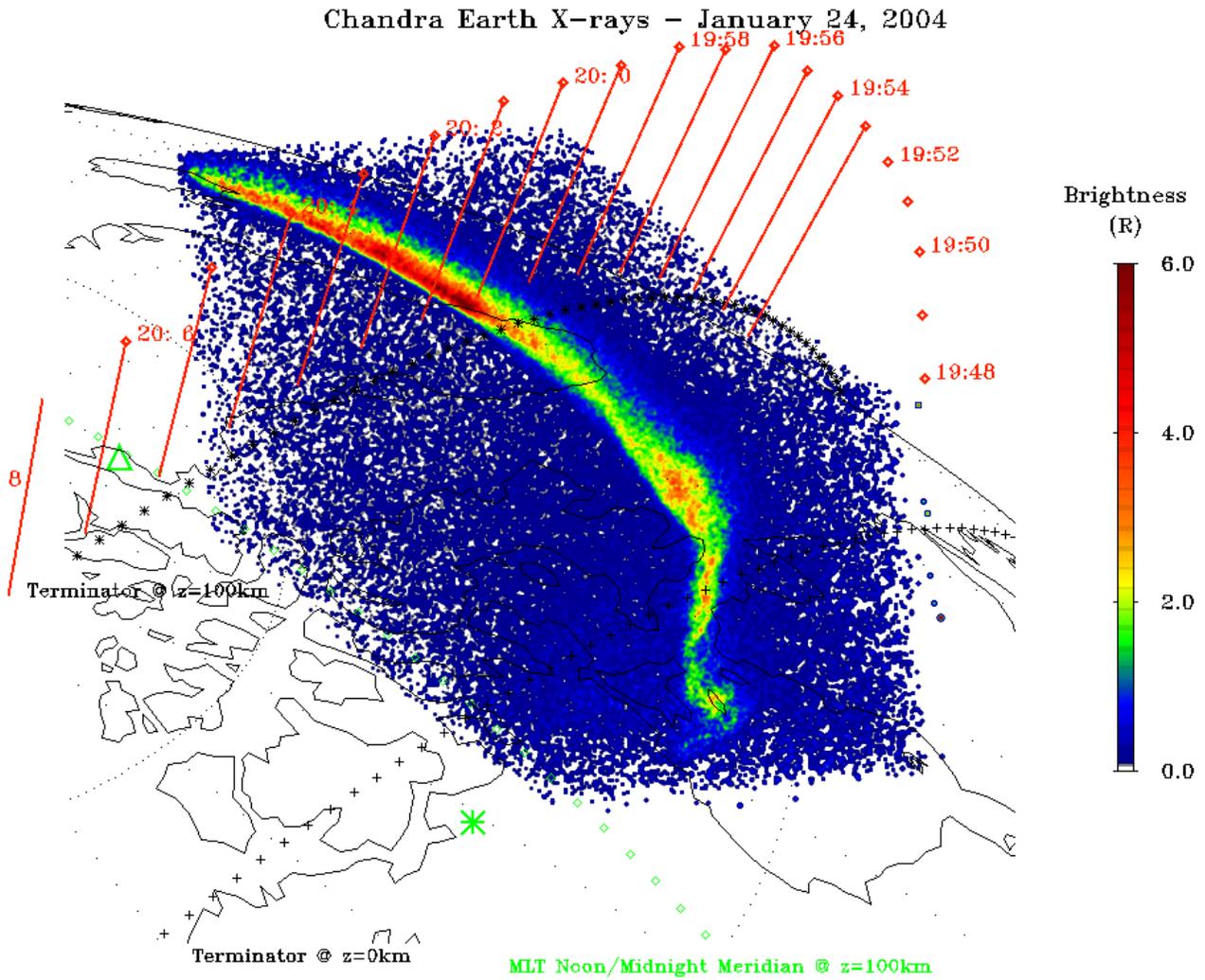

**Figure 2.** Chandra HRC-I X-ray image of auroral region on January 24, 2004 showing a bright arc. The orbital location of satellite DMSP F13 is shown by red diamonds, with 2-minute time ticks and vertical lines extending down to an altitude of 100 km. Other symbols and descriptions are as in Figure 1.

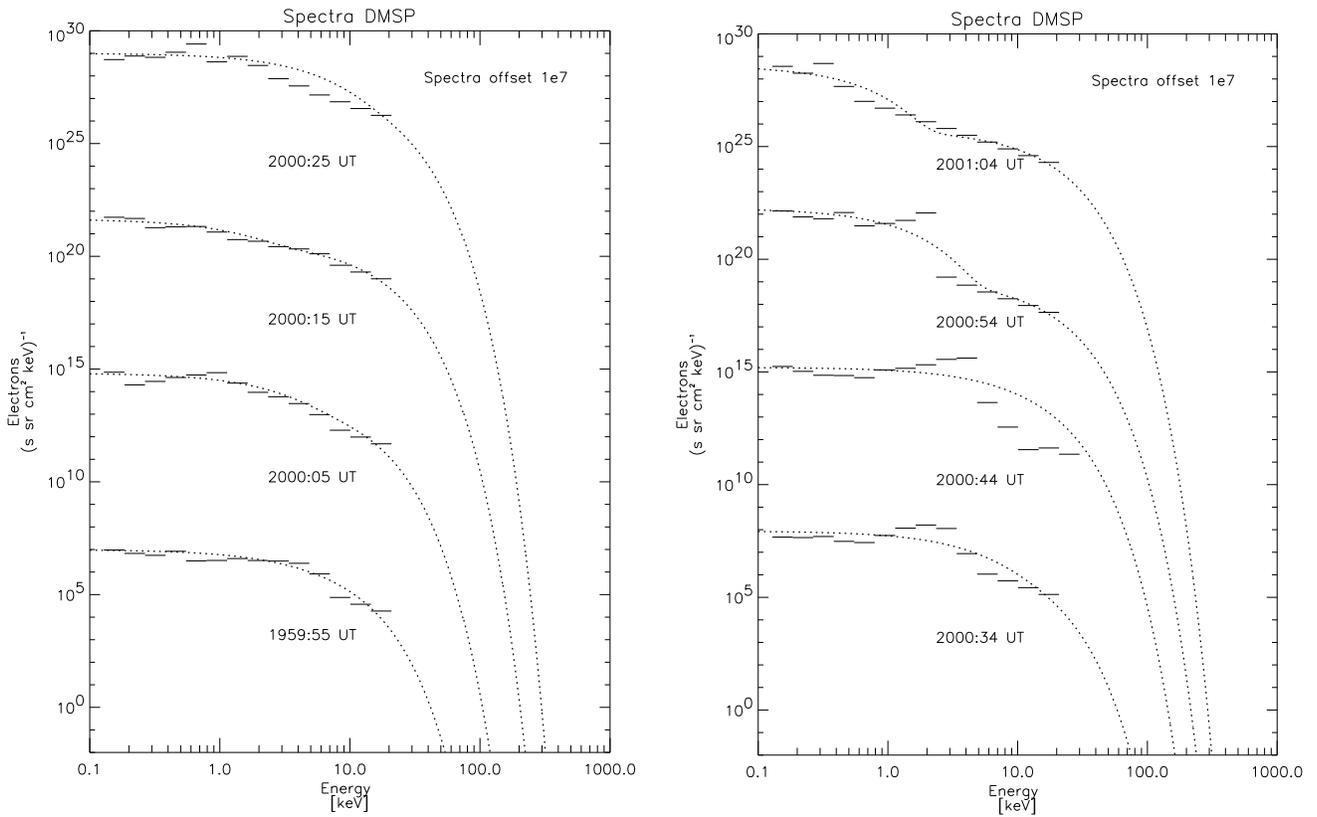

**Figure 3.** The DMSP F13 SSJ/4 measured electron spectra on January 24, 2004, showing 10-s averages, from 19:59:55 to 20:00:25 UT (left panel) and 20:00:34 to 20:01:04 UT (right panel). Horizontal lines show the DMSP measurements and dotted lines show the double exponential fit to the measurements. The spectra for 19:59:55 (left panel) and 20:00:34 (right panel) are plotted on actual scale, while other spectra are shown after shifting each spectra cumulatively by 7 orders of magnitude.

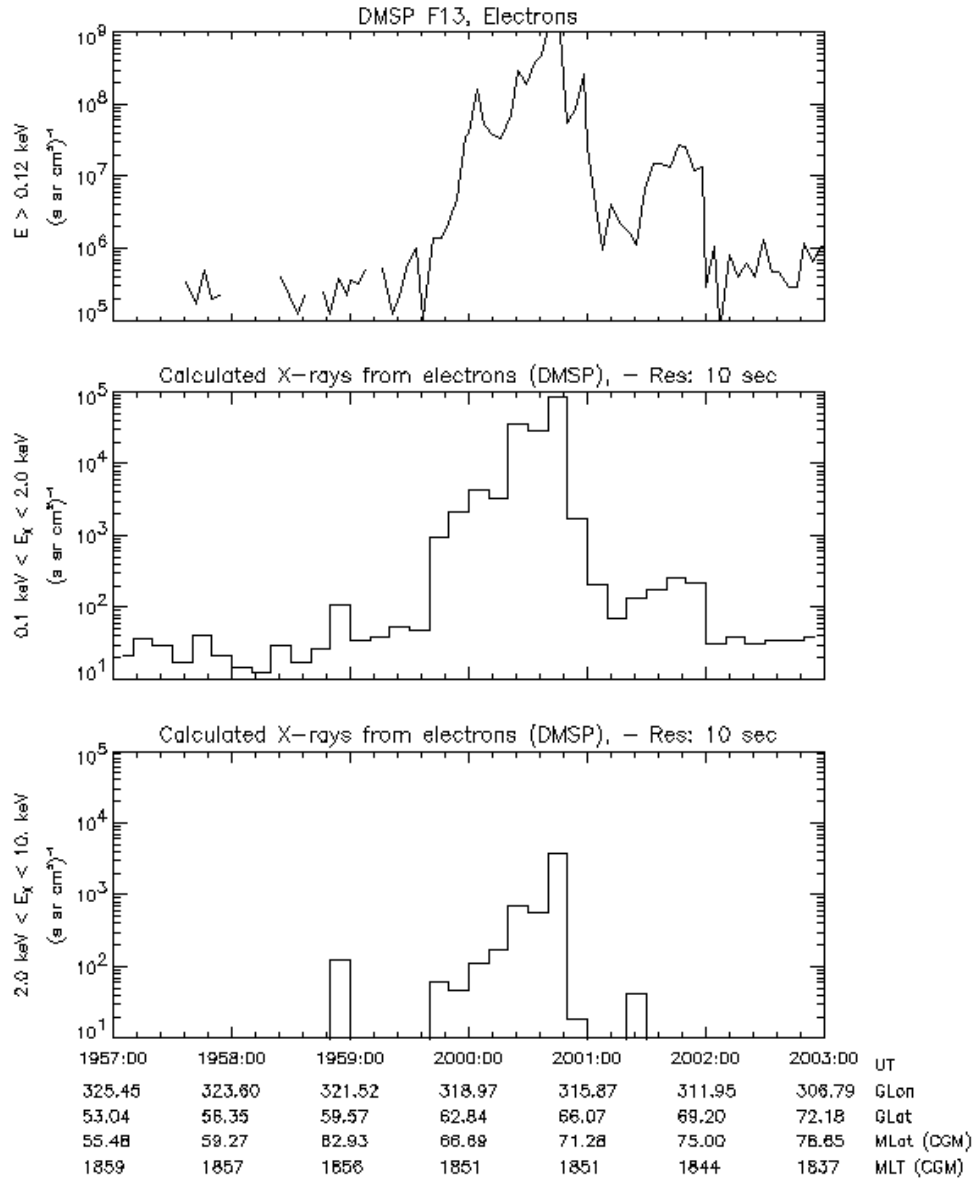

**Figure 4.** Time series of the DMSP-measured electron flux between 19:57:00 and 20:03:00 UT on 24 January 2004 (top panel) at 10-s time resolution. Calculated vertically-emitted X-ray intensity (bremsstrahlung and K-α emission lines from O (0.524 keV) and N (0.393 keV)) due to the precipitation of DMSP-measured electron flux in the energy range 0.1–2.0 keV (middle panel) and 2.0–10.0 keV (bottom panel).